# Novel hermetically sealed device to realize unconventional phonon blockade at near-micron dimensions and milli Kelvin temperatures

Jayant K. Nema[1], Srijan Gupta[2], Riya Thakkar[2] and Prabhu Rajagopal[1]

[1] *Centre for Nondestructive Evaluation and Department of Mechanical Engineering, Indian Institute of Technology Madras, Chennai, 600036, India.*

[2] *Department of Physics, Indian Institute of Technology Madras, Chennai, 600036, India.*

**Abstract**

We propose a novel design for a hermetically sealable device consisting of charged linear and nonlinear membranes driven in the Gigahertz range in vacuum setting, as a source of antibunched single phonons. Constraints for effecting phonon antibunching are found using the stationary Liouville-von Neumann master equation. Using analytical calculations, material and geometry optimization we show that sizes of the proposed system can be upscaled to near-micrometer range, in a trade-off with the system operating temperature. The results are significant to realize quantum Phononics which has much promise as a modality for sensing and computing applications.

## I. Introduction

Phononics is a relatively new branch of science and engineering encompassing the study and application of various mechanical/elastic wave phenomena (including vibration, acoustics, ultrasonics, hypersonics and thermal transport). Phonons refer to quantized states of vibration (in analogy to photons that similarly quantify light) which underlies all elastic wave phenomena [1,2]. Today elastic waves are vital to a range of applications for sensing, imaging and diagnostics in engineering and biomedicine [3-5]. Due to the generally slower propagation velocities involved, elastic wave approaches for diagnostics suffer from much poorer resolution as compared what is achievable using electromagnetics. However, the longer time scales, greater depth of penetration in various media, and non-irradiative and cost-effective transduction make elastic waves attractive for sensing and device applications. Phononics has made impressive contributions in recent years, including fundamental advances for sensing, imaging, control, vibration damping, cloaking and wavefront manipulation, and exciting phenomena such as topological and edge states [6-20].

Phononics can access a wider class of materials than its electronic or photonic counterparts. Moreover, as vibration is a universal phenomenon underlying many physical processes, Phononic systems offer the potential to revolutionize sensing and computing with recycling of waste heat and energy and on-demand or 'point-of-care' solutions. The prospect of PnC and metamaterial based novel media that can perform passive sensing, imaging and computing is exciting. However, in order to truly unveil an era of Phononics rivalling those of electronics and photonics, we would need like in those counterparts, a true source of Single or entangled Phonons, and this has not yet been experimentally realized anywhere in the world. The first observation of quantum states of vibration was made less than a decade ago through cooling to the ground state [21]. Most proposals for Phonon sources till date remain theoretical.

With the world racing in the quest for 'quantum supremacy', increasingly many researchers have also come to view Quantum Phononics as a natural base for advanced computing processes[22-23]. This is especially germane in view of the increasingly escalating demand for cooling in current sensing and computing architectures that rely on electronics or more recently, photonics. Indeed, prohibitive costs of cooling may eventually put a hold on 'Moores Law' expansion of device capabilities [24-25]. Quantum Phononics also provides the opportunity to achieve low-noise (sub-Poissonian where number distribution is such that the variance is less than the mean) imaging at very high frequencies, harnessing phenomena such as entanglement[26] and squeezed states, yielding potentially unknown precision in non-invasive materials diagnostics.

With multiple unique properties of engineering interest and scalability, phononics offers a radically new route for quantum computing[25]. These include ability to be maintained for a long time before being eventually damped, and to be able to interact with a wide range of quantum systems, such as electric, magnetic, and optical systems, making phonons promising candidates for quantum devices [22-25,27-28]. Photons, though currently the primary candidates for quantum devices, require sophisticated set-ups making practical commercial scalability challenging [29]. With a natural ability to harness waste heat and vibration, Phononic computing and sensing would also link back to the earliest era of computing and devices that involved macro-scale mechanical elements such as valves and gears.

Systems with discreet frequency-selective energy levels could be thought of as analogues of or imitate the behavior of atoms, with non-integer excitations causing release of particles (e.g., electrons in real atoms). However, in real atoms which host Fermionic (or physical/matter) particles such as electrons, multiple particles do not co-exist in the same quantum state. Thus, a source of quasi-particles such as single Phonons must have 'Antibunching', a quantum phenomenon whereby only a single quasi-particle can exist in a given system [30-31], and additional energy input to the system results in release of excess particles (phonon or photon emission).

Phonon blockade, analogous to 'photon blockade' for photons and Coulomb blockade for electrons, is typically used to effect antibunching. Initial research in this direction, deriving from similar concepts in Photonics, explored the use of non-linear oscillators (esp. micro-nanoscale beams) to achieve antibunching [32-38]. Such 'conventional' phonon blockade involves a strong non-linearity in the mechanical resonator to effect antibunching [39-42] However, the typical intrinsic nonlinearity of most micro/nanomechanical resonators is usually very weak [43-48] which makes this method difficult to implement in practice. Unconventional phonon blockade (UPB) solves this by effecting antibunching even with a weak non-linearity in a system of coupled mechanical oscillators [49].

However, the devices proposed so far require state-of-the-art nanofabrication and refrigeration capabilities, making them out impracticable for scalable practical deployment. These requirements, of manufacturability and temperature control, are stringent because unlike their photonic counterparts, because of the underlying processes, the thermal phonon number is not negligible in mechanical resonators and therefore significantly influences the phonon blockade even at temperatures of the order of several milli-Kelvins [38,49-50]. Moreover, concepts such as those in [49] are not hermetically sealable, and thus fabrication in phononic devices is a challenge.

To address the need of accessing antibunching phenomenon in larger, hotter systems we propose here, a compact, hermetically sealable device, with the goal of achieving optimal parameters of near-micrometer dimensions and near-Kelvin operation. In what follows, firstly the proposed model is described, along with equations governing the system. Results on antibunching obtainable using the proposed system, and optimization of parameters to achieve this at higher dimensions and temperatures are then discussed. The paper concludes with directions for further work.

**II. Model: Compact, hermetically sealable device to realize anti-bunching in phonons**

As shown in Fig. 1 below, the proposed system consists of two Coulomb-coupled, circumferentially clamped, identical circular membranes under pre-tension(s) that are given an equal and opposite charge q, and are actuated using time-varying electric fields. The casing is rigid & fixed in space to prevent boundary movement. This system is hermetically sealable, and hence yields a convenient route for large-scale device fabrication and integration within circuits.

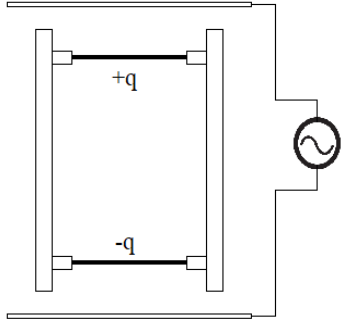

*Fig 1. Schematic of the proposed device concept proposed to achieve phonon antibunching*

The driving frequency, material properties of the membranes, quantity of charge, separation and pre-tension imparted are thus the tunable parameters for the device. We assume that the resonator 1 is linear and is harmonically driven externally by force amplitude $F$ and frequency $\omega_d$. The Hamiltonian for this system can be written (assuming terms to be divided by $\hbar$):

$$H = \omega_1 \hat{b}_1^\dagger \hat{b}_1 + \omega_2 \hat{b}_2^\dagger \hat{b}_2 + J(\hat{b}_1^\dagger + \hat{b}_1)(\hat{b}_2^\dagger + \hat{b}_2) + F(\hat{b}_1^\dagger e^{i\omega_d t} + \hat{b}_1 e^{-i\omega_d t}) + U(\hat{b}_2^\dagger + \hat{b}_2)^4 \quad (1)$$

where $\hat{b}_j(\hat{b}_j^\dagger)$ is the annihilation (creation) operator for the phonon mode of the j-th mechanical resonator with resonance frequency $\omega_j$ and damp rate $\gamma_j$ ($j = 1,2$). Coupling strength between the two mechanical resonators is denoted by $J$, and $U$ is the nonlinearity of mechanical resonator 2. Pre-tension and material of the membranes dictate resonance frequency while charge and separation parameters control the coupling strength $J$ by the following relation (see Appendix for derivation):

$$J = -\frac{1}{2}\frac{d^2(U_e(z))}{dz^2}\sqrt{\frac{1}{m_1 m_2 \omega_m^2}} \quad (2)$$

where $m_1, m_2$ are the masses and $\omega_m$ the frequency of resonators (ie. $\omega_m = \omega_0$). In order to evaluate the double derivative in Equation (2), we write an expression for the potential energy $U_e$ between two uniformly charged discs of same radius and separation of z [51] as

$$U_e(z) = \frac{4k_e Q_1 Q_2}{R}\left(-\frac{a}{2} + \frac{a}{6\pi}\times\left[(4-a^2)E\left(-\frac{4}{a^2}\right) + (4+a^2)K\left(-\frac{4}{a^2}\right)\right]\right) \quad (3)$$

where $a = |z|/R$ For real values of an argument m, $K(m)$ and $E(m)$ are complete elliptic integrals of first and second kind. Using (1)&(2) we can find out the value of $J$ for any value of charge and separation and tune it by changing the mass values. For simplicity and without loss of physics, in the following, we assume that the mechanical resonators share the same frequency and decay rate, and that the coupling strength $J$ and the nonlinearity $U$ are much smaller than resonance frequency $\omega_d$, i.e., $\omega_1 = \omega_2 = \omega_0$; $\gamma_1 = \gamma_2 = \gamma$; $J \ll \omega_d$; $U \ll \omega_d$; Under these assumptions, we can neglect anti-rotating wave terms rewrite the Hamiltonian as

$$H = \Delta(\hat{b}_1^\dagger \hat{b}_1 + \hat{b}_2^\dagger \hat{b}_2) + J(\hat{b}_1^\dagger \hat{b}_2 + \hat{b}_2^\dagger \hat{b}_1) + F(\hat{b}_1^\dagger + \hat{b}_1) + U\hat{b}_2^\dagger \hat{b}_2 \quad (4)$$

where $\Delta$ is the detuning of the mechanical resonance frequency from the driving frequency.

The Hamiltonian in Eq. (4) describes a model of driven-dissipative coupled nonlinear mechanical resonators, which is mathematically similar to the optical counterpart [52]. Unlike photon blockades, temperature has significant influence on phonon correlation owing to the low energy of individual phonons. With inclusion of temperature factor, Liouville-von Neumann master equation for the density matrix gives

$$\frac{d\hat{\rho}}{dt} = \hat{L}\hat{\rho} = -i[H,\hat{\rho}] + \sum_{n=1,2}\frac{\gamma}{2}\{(n_{th}+1)D[\hat{b}_n]\hat{\rho} + n_{th}D[\hat{b}_n^\dagger]\hat{\rho}\} \quad (5)$$

with Lindblad operator $D[\hat{A}]\hat{\rho} = 2\hat{A}\hat{\rho}\hat{A}^\dagger - \hat{A}^\dagger\hat{A}\hat{\rho} - \hat{\rho}\hat{A}^\dagger\hat{A}$. Here, $n_{th} = (\exp\left(\frac{T_0}{T}\right)-1)^{-1}$ is the average phonon number of the mechanical resonators at the temperature T (we assume that the temperatures of two nanomechanical beams are the same) with $T_0 = \hbar\omega/K_B$. Here, we have neglected the pure dephasing of the resonators because the dephasing rates are usually much smaller than other decay rates[53]. The phonon states for mechanical resonator 1 are described by the equal-time second-order correlation function:

$$g^{(2)}(0) = \frac{Tr(\hat{b}_1^\dagger \hat{b}_1^\dagger \hat{b}_1 \hat{b}_1 \hat{\rho}_{ss})}{Tr(\hat{b}_1^\dagger \hat{b}_1 \hat{\rho}_{ss})^2} \quad (6)$$

where $\hat{\rho}_{ss}$ is the steady-state density matrix by setting $\frac{d\hat{\rho}}{dt} = 0$ in eq. (3). and density matrix is

$$\hat{\rho} = \sum_{m,n=0}^{N}\rho_{mn,m',n'}|mn\rangle\langle m'n'| \quad (7)$$

on the basis of phonon number states $|mn\rangle$, where $m$ denotes the phonon number in mechanical mode 1 and $n$ denotes the phonon number in mechanical mode 2. To find the steady-state density matrix, we need to find the eigenmatrix $\hat{\rho}_{ss}$ of superoperator $\hat{L}$ when it has an eigenvalue equal to 0, i.e., $\hat{L}\hat{\rho} = \lambda\hat{\rho} = 0 \rightarrow \lambda = 0$ Such an eigenvalue problem can be numerically solved using the method in ref. [54]. In the numerical calculation, we set the phonon number n = 10, m = 10, which is sufficiently large to ensure the convergence of the results in this work.

## III. Results

It has been shown that $0.04T_0$ (where $T_0 = h\omega_0/2\pi K_B$) is a suitable system temperature for phonon antibunching [49]. As the temperature is pushed higher, antibunching fades. This puts an upper bound on our system temperature. Following design in [55] conceives a compact 3mK refrigeration in latter of 1980s, with modern day machinery this threshold of cooling capacity should be feasible, hence this has been chosen as the lower bound for our system temperature.

$$3\text{mK} \leq T_{sys} \leq 0.04T_0 \tag{8}$$

The smallest system dimension ($z_{sys}$) and the system temperature ($T_{sys}$), are the two objectives to be maximized. We take the largest allowed value for system temperature ie., ($T_{sys} = 0.04T_0$), and express $T_0$ in terms of system's natural frequency

$$T_{sys} = f_{sys} \times \beta; \quad \beta = \frac{h}{K_B} = (1.9196 \times 10^{-12} \text{ Kelvin seconds}) \tag{9}$$

$f_{sys}$ can be related to ($z_{sys}$) using structural mechanics of resonator[56]. Using (9) $f_{sys}$ can be written in terms of $T_{sys}$ giving us relations between $T_{sys}$ & $Z_{sys}$. With simple algebraic manipulation it is found that for all the geometries studied, trade-off curves between $T_{sys}$ & $Z_{sys}$ take the following form

$$T_{sys} \times Z_{sys} = P \tag{10}$$

This means that for a given value of $T_{sys}$, maximizing $Z_{sys}$ would mean maximizing $P$ and vice versa. The structural mechanics expression of $P$ depends on the geometric configuration and material factors pertaining to resonator design, which when grouped and re-written gives

$$P = \alpha_g \times P_a \times P_g \times P_m \times \beta \tag{11}$$

Here $\alpha_g, P_a$ & $P_g$ are dimensionless and configuration (geometry) dependent. $\alpha_g$ is a numerical value dependent on the configuration chosen and $P_a, P_g$ & $P_m$ are performance parameters that can be tuned by changing the aspect ratio, cross-sectional geometry and material properties respectively. $P_m$ & $P$ have dimensions of [L][T$^{-1}$] and [L][K] respectively.

Once the geometry configuration is chosen $\alpha_g$ becomes determined and our optimization problem splits into three sub-optimization problems pertaining to optimization of each of the individual performance metrics $P_a, P_g$ & $P_m$. For ideally compliant structures the material performance metric $P_m$ becomes the square root of specific strength $\sqrt{\sigma/\rho}$, and for ideally elastic structures it becomes the square root of specific stiffness $\sqrt{E/\rho}$, where $\sigma, \rho$ & $E$ are of the material chosen for linear resonator.

In this study, the compliant structure studied is the circumferentially clamped isotropic circular membrane and elastic structures studied are rectangular cross-section beams (both solid & hollow), circular cross-section beams (both solid & hollow), and I-beams. Expressions for $\alpha_g, P_a, P_g$ & $P_m$ along with their optimal values for these are listed in the Table 1. The fraction of material removed radius-wise, height-wise & width-wise are $r, c$ & $d$ respectively (Evaluation performed at r = 0.8, c = 0.75, d = 0.8). With regards to material metric $P_m$ it is found that in circular membranes, Graphene optimizes the value of $P_m$ to 7900 m/s[57]; and in beams, diamond optimizes the value of $P_m$ to 18600m/s[58]. Maximal value of $t/R$ = 0.2 (for thin disc approximation), $H/L$ & $D/L$ = 0.1 (for slender beam approximation)

| Resonator Geometry | $\alpha_g$ | $P_a$ | $P_g$ | $P_m$ | $P_a$ optimal | $P_g$ optimal | P(m-K) optimal |
|---|---|---|---|---|---|---|---|
| Circular Membrane | $\frac{2.405}{2\pi}$ | $\frac{t}{R}$ | $\frac{1}{1}$ | $\sqrt{\sigma/\rho}$ | 0.2 | 1 | $9.47 \times 10^{-10}$ |
| Hollow Rectangular Beam | $\frac{22.373}{2\pi}$ | $\frac{H^2}{L^2}$ | $\frac{1}{\sqrt{12}}\sqrt{\frac{1-c^3d}{1-cd}}$ | $\sqrt{E/\rho}$ | 0.01 | 0.371 | $4.72 \times 10^{-10}$ |
| Hollow Circular Beam | $\frac{22.373}{2\pi}$ | $\frac{D^2}{L^2}$ | $\frac{1}{\sqrt{16}}\sqrt{1+r^2}$ | $\sqrt{E/\rho}$ | 0.01 | 0.320 | $4.06 \times 10^{-10}$ |
| Solid Rectangular Beam | $\frac{22.373}{2\pi}$ | $\frac{H^2}{L^2}$ | $\frac{1}{\sqrt{12}}$ | $\sqrt{E/\rho}$ | 0.01 | 0.288 | $3.66 \times 10^{-10}$ |
| Solid Circular Beam | $\frac{22.373}{2\pi}$ | $\frac{D^2}{L^2}$ | $\frac{1}{\sqrt{16}}$ | $\sqrt{E/\rho}$ | 0.01 | 0.250 | $3.17 \times 10^{-10}$ |
| I-Beam | $\frac{22.373}{2\pi}$ | $\frac{H^2}{L^2}$ | $\frac{1}{\sqrt{12}}\sqrt{\frac{1-c^3+c^3d}{1-c+cd}}$ | $\sqrt{E/\rho}$ | 0.01 | 0.371 | $4.72 \times 10^{-10}$ |

*Table 1. Performance metrics and their optimal values for different geometries,*

## IV. Discussion

Trade-off curves between $T_{sys}$ & $Z_{sys}$ are plotted in Fig. 2, while the results of optimizations of $Z_{sys}$ and $T_{sys}$ values pertaining to each of the configurations are listed in Tables 2 and 3.

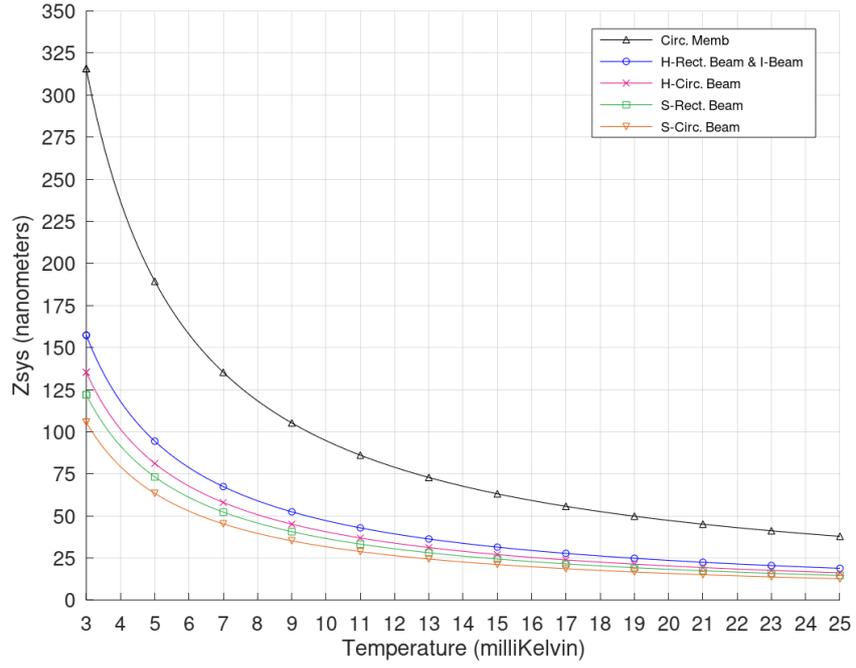

*Fig 2. Trade-off curves between $T_{sys}$ and $Z_{sys}$ for different geometries as discussed in Table*

| Configuration | $Z_{sys}$ for 3mK | $Z_{sys}$ for 25mK |
|---|---|---|
| Circular Membrane | 315 nm | 37.9 nm |
| H-Rectangular Beam | 157 nm | 18.8 nm |
| H-Circular Beam | 135 nm | 16.2 nm |
| S-Rectangular Beam | 122 nm | 14.6 nm |
| S-Circular Beam | 105 nm | 12.7 nm |
| I-Beam | 157 nm | 18.8 nm |

*Table 2. $Z_{sys}$ values for different configurations at different $T_{sys}$ values.*

| Configuration | $T_{sys}$ for 0.1 $\mu$m | $Z_{sys}$ for 25mK |
|---|---|---|
| Circular Membrane | 9.47 mK | 37.9 nm |
| H-Rectangular Beam | 4.72 mK | 18.8 nm |
| H-Circular Beam | 4.06 mK | 16.2 nm |
| S-Rectangular Beam | 3.66 mK | 14.6 nm |
| S-Circular Beam | 3.17 mK | 12.7 nm |
| I-Beam | 4.72 mK | 18.8 nm |

*Table 3. $T_{sys}$ values for different configurations at different $Z_{sys}$ values.*

For the circular membrane, the *J* value is tunable and *F* value is adjusted externally. Parametric sensitivity analysis is carried out to check effect of non-linearity *U*, coupling strength *J* and forcing *F* on the second order correlation function $g^{(2)}(0)$. The resulting variations are plotted in Fig 2., The results indicate that for J* = 110, U* = 3 × $10^{-5}$, F* = 10, Δ* = 0.29, $g^{(2)}(0)$ << 1 which implies strong anti-bunching effect. Note that J* = $J/\gamma$, likewise U*, F* and Δ* are also non-dimensional analogues generated by dividing with decay rate ($\gamma$).

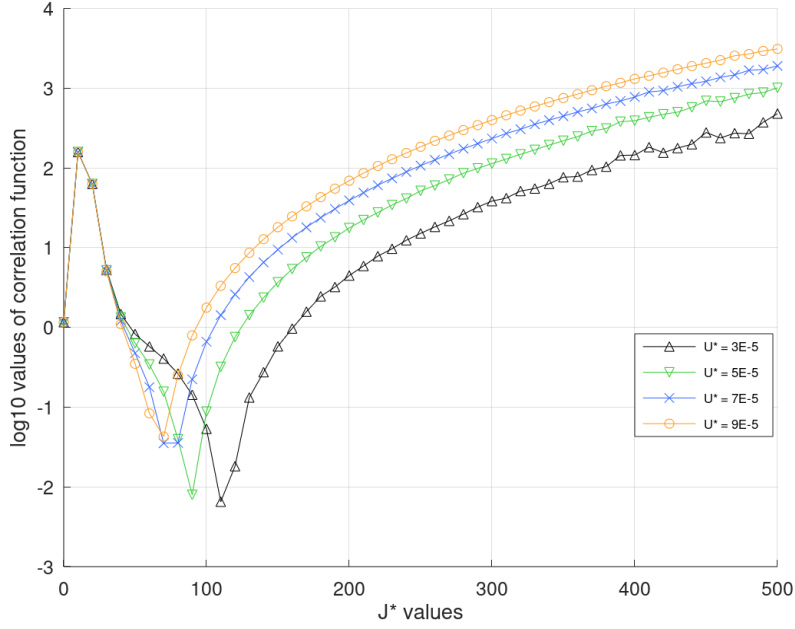

*Fig 2. Variation of $g^{(2)}(0)$ with coupling strength; Here, $F = 10\gamma$ and $\Delta = 0.29\gamma$*

## V. Conclusions and further work

In this work we have shown that if the material properties are also optimized by using Diamond for beams, hollow rectangular beam (with central 75% height-wise and 80% width-wise material removed ie. c=0.75, d=0.8) the dimensions can be pushed up to 157nm, pushing its size into the near-micro regime. If Graphene can be used for membrane we can scale further to 315nm at 3mK. Authors hope this work contributes towards making phonon-based quantum devices more accessible and scalable. Furthermore, for all the geometries considered the results show temperature requirement at 0.1 micron scale is greater than 3mK hence above the set lower-bound. Further and ongoing work at our group is studying the development of density of states from the Hamiltonian and hence explore the origins of antibunching in nanomechanical resonator systems. Approaches to fabricate and test the device proposed here, are also being initiated.

**Data Availability**

The data that support the findings of this study are available from the corresponding author upon reasonable request.

**Supplementary Material**

The Supplementary Material provides a detailed derivation of the coupling strength J as given in Equation 2, based on the procedure in Ref. [29]

# Appendix

Following the procedure in ref.[29] we deduce a general expression for coupling strength J in terms of model parameters. We shall denote the generalized equations with lettered numbers (1g)-(5g) and their corresponding applied forms from ref [29] with lettered numbers (1a)-(5a). To begin the derivation we first write the original equation for coulombic coupling interaction

$$H_{co} = J(\hat{b}_1^\dagger + \hat{b}_1)(\hat{b}_2^\dagger + \hat{b}_2) \qquad (1g)$$

Hamiltonian is then written in terms of coulombic interaction energy $U_e$ where $x_1$ and $x_2$ are displacements from equilibrium positions.

$$H_{co} = U_e(d + x_1 - x_2) \qquad (2g)$$

$$H_{co} = \frac{k_e q_1 q_2}{|d + x_1 - x_2|} \qquad (2a)$$

The expression for Hamiltonian is expanded in Taylor form till quadratic term. Note- Here, $f'(t)$ & $f''(t)$ will be denoting 1st & 2nd derivatives of $f(t)$ respectively

$$H_{co} = U_e(d) + U_e'(d) \cdot (x_1 - x_2) + \left(\frac{1}{2}\right) U_e''(d) \cdot (x_1 - x_2)^2 \qquad (3g)$$

$$H_{co} = \frac{k_e q_1 q_2}{d}\left(1 - \frac{x_1 - x_2}{d} + \left(\frac{x_1 - x_2}{d}\right)^2\right) \qquad (3a)$$

Here, the first term is a constant term and the second one is a linear term which can be absorbed into the definition of the equilibrium positions. The last term consists of two parts: one part refers to the small frequency shift of the original frequencies and can be neglected by re-normalising the mechanical frequencies, and the other part is the coupling term between the oscillators [29]. Therefore, we obtain the Coulomb interaction between the mechanical oscillators as

$$H_{co} = -\frac{U_e''(d)}{2}(-2x_1 x_2) \qquad (4g)$$

$$H_{co} = \left[\frac{k_e q_1 q_2}{d^3}\right](-2x_1 x_2) \qquad (4a)$$

From [29] we also have

$$J = \left[\frac{k_e q_1 q_2}{d^3}\right]\sqrt{\frac{1}{m_1 m_2 \omega_m^2}} \qquad (5a)$$

Substituting $\left[\frac{k_e q_1 q_2}{d^3}\right]$ from Eqn.(5a) into Eqn.(4a) we get

$$H_{co} = J \cdot \sqrt{m_1 m_2 \omega_m^2} \cdot (-2x_1 x_2)$$

Comparing this with Eqn.(4g) gives us the required general expression

$$J = -\frac{1}{2}U_e''(d)\sqrt{\frac{1}{m_1 m_2 \omega_m^2}} \qquad (5g)$$

This completes the derivation of formula (2) used in our work.